\documentstyle[aps,prl,epsf,floats]{revtex}
\begin{document}
\twocolumn[\hsize\textwidth\columnwidth\hsize\csname@twocolumnfalse%
\endcsname


\title{ Orbital excitations in LaMnO$_3$}
\author { Jeroen van den Brink}
\address{ Computational Materials Science and Mesa$^+$ Institute, 
Faculty of Applied Physics,\\
University of Twente, P.O. Box 217, 7500 AE Enschede, The Netherlands}

\date{\today}
\maketitle

\begin{abstract}
We study the recently observed orbital excitations, orbitons, and treat electron-electron 
correlations and lattice dynamics on equal footing. It is shown that the orbiton energy and 
dispersion are determined by both correlations and lattice-vibrations.  
The electron-phonon coupling causes satellite structures in the orbiton spectral function and
the elementary excitations of the system are mixed modes with both orbital and phonon character. 
It is proposed that the satellite structures observed in recent Raman-scattering experiments 
on LaMnO$_3$ are actually orbiton derived satellites in the phonon spectral function, 
caused by the phonon-orbiton interaction.
\end{abstract}

]


Elementary properties of electrons in atoms and solids are determined by their charge and spin, but in many correlated electron materials also by their orbital degree of freedom~\cite{Imada98,KK73}. 
In a free transition metal atom, with an open $d$ shell, the $3d$ energy levels are
five-fold orbitally degenerate, where each orbital state corresponds to a different quadrupolar charge-distribution in real-space. In correlated Mott-insulators, among which there are 
many transition metal oxides (TMO's), monopolar charge excitations, that involve moving charge from
one atom to another, are only possible at high energies because of the large Coulomb interaction 
between electrons. 
However, low energy multipolar charge excitations -corresponding to orbital excitations- are possible 
as they can be locally charge-neutral.
 
In a solid the orbital degeneracy of a free ion is lifted. There are two physically distinct 
mechanisms. One possibility is that the electron-electron interaction splits the states 
via the superexchange and in this way couples orbitals to the spin and relates orbital- to 
magnetic-order~\cite{KK73}. 
On the other hand, also the electron-phonon interaction, which gives rise to orbital order 
accompanied by a Jahn-Teller lattice distortion~\cite{Jahn37}, can lift the degeneracy.  
The resulting orbital order is found in many TMO's, ranging from titanates 
(e.g. YTiO$_3$~\cite{Ichikawa00}), vanadates (e.g. V$_2$O$_3$~\cite{Castellani78}, 
LiVO$_2$~\cite{Pen97}, YVO$_3$~\cite{Ren98}) and manganites 
(e.g. LaMnO$_3$~\cite{Gooden63}, Nd$_{1/2}$Sr$_{1/2}$MnO$_3$~\cite{Kuwahara95}), 
to cuprates (e.g. KCuF$_3$~\cite{KK73}). 
The quadrupolar charge ordering -orbital ordering- should give rise to elementary excitations 
with orbital signature, as the order causes a breaking of symmetry in the orbital sector. 
The existence of such excitations, orbitons, was predicted in the 70's and theoretically studied ever since~\cite{KK73,Ishihara97,BHMO99,Allen99}, but only very recently orbitons were observed for 
the first time in a Raman scattering experiment on LaMnO$_3$~\cite{Saitoh01}. 
The observed orbitons were interpreted by some as being due to electron correlations~\cite{Saitoh01}, 
but others suggest that they originate from the electron-lattice coupling~\cite{Allen01}. 
This motivates us to address the question of the origin of the orbitons, also because it is important 
to establish the precise nature of the orbitons as they, in turn, have a large effect 
on spin~\cite{Feiner97,BSKS98} and monopolar charge excitations~\cite{BHO00,Perebeinos00}.

We use a realistic model Hamiltonian for LaMnO$_3$ that incorporates both superexchange and 
electron-phonon coupling. We study the Hamiltonian first in the localized limit in order
to gain more physical insight into the problem.
This approach illustrates that for groundstate properties it is usually sufficient to treat the 
Jahn-Teller phonons as classical entities~\cite{Hotta99}, but that for orbital dynamics it is 
essential to treat the lattice-vibrations quantum-mechanically.  
The full calculation shows that in LaMnO$_3$ the orbiton has exchange and lattice character: 
its energy and dispersion are determined by both correlations and phonons.
We propose that the peaks in the Raman-scattering data on LaMnO$_3$~\cite{Saitoh01}, 
are orbiton derived satellites in the phonon spectral function, which arise due to the
mixing of the orbital and phonon modes.

{\it Hamiltonian.} We consider the two-fold degenerate manganese $e_g$ states, with one electron 
per site. The electron can either be in the $x^2-y^2$ or $3z^2-r^2$ 
orbital, or in any linear combination of these two states. 
The interaction between neighboring orbitals is mediated by the superexchange and the 
electron-phonon (e-p) interaction couples the electron to the two-fold degenerate Jahn-Teller 
phonons that have $e_g$ symmetry. We also take into account the dispersion of these phonons. 
Let us split up the Hamiltonian in an orbital, e-p 
and free phonon part: $H=H_{orb}+H_{ep}+H_{ph}$, with
\begin{eqnarray}
H_{orb}^0+H_{e-p}^0 = \sum_{\langle ij\rangle_{\Gamma}} 
J_{\Gamma} T^{\Gamma}_{i} T^{\Gamma}_{j}  +
2g \sum_{i} \tau^z_i Q_{3i} + \tau^x_i Q_{2i},
\label{H_orb0}
\end{eqnarray}
where the sum is over neighboring sites $\langle ij \rangle$  along the 
$\Gamma = a,b,c$ crystallographic axes.
The orbital operators $T^{\Gamma}$ can be expressed in terms of the 
Pauli matrices $\tau$: 
$
T^{a/b}_{i} = ( \tau^z_i \pm \sqrt{3} \tau^x_i )/2,
$
and
$
T^{c}_{i} = \tau^z_i.
$
The e-p coupling constant is denoted by $g$ and 
the phonon operators of the so-called $Q_2$ and $Q_3$ Jahn-Teller modes
with $e_g$ symmetry are
$
Q_{2/3i}=q_{2/3i}^{\dagger} + q_{2/3i}.
$
The free phonon part of the Hamiltonian is then
\begin{equation}
H_{ph}^0 = \omega_0 \sum_i \left[ q_{3i}^{\dagger}q_{3i} +
                               q_{2i}^{\dagger}q_{2i} \right]
+ \omega_1 \sum_{\langle ij\rangle_{\Gamma}} 
                Q^{\Gamma}_{i} Q^{\Gamma}_{j}, 
\label{H_ph0}
\end{equation}
where the local phonon energy is given by $\omega_0$ and the nearest
neighbor coupling between the phonons by $\omega_1$. 
The coupled Jahn-Teller modes along the three spatial axes are 
$
Q^{a/b}_{i} = (Q_{3i} \pm \sqrt{3} Q_{2i})/2
$
and
$ 
Q^{c}_{i} = Q_{3i}
$
~\cite{Khaliullin00}.
Note that the orbital excitations are locally charge neutral and do
therefore not couple to breathing mode distortions.

The Hamiltonian Eqs.(\ref{H_orb0},\ref{H_ph0}) is general and one needs to 
make it more specific in order to describe LaMnO$_3$.
This is a prototype of an orbital ordered and Jahn-Teller distorted system and 
its crystallographic structure basically consists of corner connected MnO$_6$ octahedra, where 
the space between octahedra is filled with Lanthanum atoms. 
Due to the correlation and Jahn-Teller coupling the 
octahedra are elongated, with the axis of elongation along the crystallographic $a$-direction on one 
sublattice, and along the $b$-direction on the other, such that the reciprocal lattice vector for the orbital order is ${\bf Q} = (\pi,\pi,0)$. An intersite phonon coupling arises 
because the elongation of a MnO$_6$ octahedron induces an contraction of the neighboring octahedron, and vice versa, as the octahedra have one corner in common. 
The orbital order is formally incorporated in the Hamiltonian by performing a rotation of the $T$ 
operators~\cite{BHMO99}, with equal orbital exchange constants 
$
J_a=J_b=J
$
along the two axes in the plane.

{\it Transformations.} 
In analogy with linear spin wave theory,
the orbital modes can be found by performing a Holstein-Primakov
transformation~\cite{BHMO99}. We introduce on each site $i$ the bosonic orbital 
operators $q^{\dagger}_{1i}$ and $q_{1i}$: 
$
\tau^z_i = \frac{1}{2} - q^{\dagger}_{1i} q_{1i} 
$
and
$
\tau^x_i = \frac{1}{2} ( q^{\dagger}_{1i} + q_{1i}).
$
We see from Eq.~(\ref{H_orb}) that this transformation introduces
a term that is linear in the phonon mode $Q_3$ in $H_{e-p}$. This is a
consequence of the long range orbital order that we assumed to be present
from the beginning: the lattice deforms according to the symmetry of
the occupied orbital on each site. The linear term can be gauged away
by introducing 
$
q_3 \rightarrow q_3 + \eta,
$
where the shift $\eta$ is given by
$
\eta=g/(\omega_0-6 \omega_1).
$
After this shift we collect the quadratic and cubic terms
in the bosonic operators and find in Fourier space
\begin{eqnarray}
H_{orb} &=& \sum_{k} 
(3J +4 g \eta) q^{\dagger}_{1k} q_{1k} 
- \frac{J \gamma_{1k}}{4} Q_{1k} Q_{1-k} \label{H_orb} \\
H_{e-p} &=& g\sum_{k,q} 
2 \  q^{\dagger}_{1 k-q} q_{1k} Q_{3q} + Q_{1k}  Q_{2-k} \label{H_e-p}  \\
H_{ph} &=& 
\omega_0 \sum_k 
\left[ q_{3k}^{\dagger}q_{3k}+q_{2k}^{\dagger}q_{2k} \right. \nonumber \\
&+& 
\left. 
\frac{\omega_1}{2} (
 \gamma_{2 k} Q_{2k}Q_{2-k}
 -3 \gamma_{\parallel k} Q_{3k}Q_{3-k} ) \right],
\label{H_ph}
\end{eqnarray}
with
$
Q_{\nu k} =  q^{\dagger}_{\nu k} + q_{\nu -k} 
$
and the dispersions
$
\gamma_{1k} =  
\gamma_{\parallel k} + \gamma_{\perp k} J_c / J,   
$
$
\gamma_{2k} = 2\gamma_{\perp k} - \gamma_{\parallel k},
$
where 
$
\gamma_{\parallel k} = (\cos k_x + \cos k_y)/2
$
and
$
\gamma_{\perp k} = \cos k_z
$
\cite{mixed_terms}.

Three important consequences of the orbiton-phonon coupling are
present in  Eqs.~(\ref{H_orb}-\ref{H_ph}). First, the coupling to the lattice
moves the orbiton to higher energy an amount $4g\eta$. 
This shift has a straightforward physically meaning:
it is phonon contribution to the crystal-field splitting of the $e_g$ states
caused by the static Jahn-Teller lattice deformation~\cite{Bala00}.
If, however, an orbital excitation is made, it strongly interacts
with the $Q_3$ phonon (Eq.~\ref{H_e-p}), so that the orbital excitation 
can be dynamically screened by the Jahn-Teller phonons and lowered in energy.
The crystal-field splitting and screening are strongly competing as
both are both governed by the energy scale set by the e-p coupling. 
Finally, the orbital and $Q_2$ phonon modes mix, 
as is clear from the second term of $H_{e-p}$. This implies that
the true eigenmodes of the coupled orbital-phonon system have both 
orbital and phonon character. 

\begin{figure}
\epsfxsize=80mm
\centerline{\epsffile{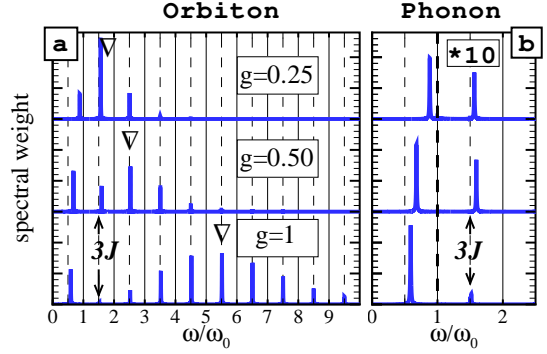}}
\vspace*{2mm}
\caption{
(a) Orbiton (Q$_1$) and (b) phonon (Q$_2$) spectral function in the localized limit. The first
pole due to the orbital exchange is indicated at $3J$. In the orbital spectrum the static crystal-field 
energy $\bar{J}$ is indicated by $\nabla$. The spectral weight in (b) is multiplied by 10 
for $\omega > \omega_0$.
}
\label{Fig:local}
\end{figure}

{\it Localized limit.} We illustrate the three physical effects of the orbiton-phonon coupling, 
discussed above, by considering the Hamiltonian of Eqs.~(\ref{H_orb}-\ref{H_ph}) first in the 
localized limit, neglecting all dispersion. The Hamiltonian then reduces to
\begin{eqnarray}
H_{loc} =&
(\bar{J} + 2g Q_{3} ) q^{\dagger}_{1} q_{1} + 
\omega_0 (q_{3}^{\dagger}q_{3}+q_{2}^{\dagger}q_{2}) 
+ g Q_{1}  Q_{2}, 
\label{H_loc}
\end{eqnarray}
with
$
\bar{J}= 3J+4g \eta,
$
which is the sum of the local orbital exchange energy and static phonon contribution
to the crystal-field splitting.
The Hamiltonian without the last term is exactly solvable by a 
canonical transformation~\cite{Mahan} so that we can obtain the expressions
for the 6 by 6 matrix of bosonic Green's functions 	
$
D_{11}(\nu,t-t^{\prime}) = -i \langle |T q^{\dagger}_{\nu}(t) 
q_{\nu}(t^{\prime}) | \rangle,
$
$
D_{12}(\nu,t-t^{\prime}) = -i \langle |T q_{\nu}(t) 
q_{\nu}(t^{\prime}) | \rangle,
$
$
D_{21}(\nu,t-t^{\prime}) = -i \langle |T q^{\dagger}_{\nu}(t) 
q^{\dagger}_{\nu}(t^{\prime}) | \rangle,
$
$
D_{22}(\nu,t-t^{\prime}) = -i \langle |T q_{\nu}(t) 
q^{\dagger}_{\nu}(t^{\prime}) | \rangle,
$
with $\nu=1,2,3$.
The last term in Eq.~(\ref{H_loc}) couples the orbiton and $Q_2$ phonon mode and 
introduces a 
self energy~\cite{ToBePub}. 
We use the on-site Jahn-Teller vibrational energy $\omega_0= 80$ meV 
as unit of energy, and $J=\omega_0/2$, in accordance with Refs.~\cite{Saitoh01,BHO00}.
In Fig.~\ref{Fig:local}a 
the calculated orbiton spectral function, $-\frac{1}{\pi}Im \ D(1,\omega)$, is
plotted. For small e-p coupling $g$, most of the spectral weight is in the pole at 
$\omega \approx 3J$, and phonon satellites with decreasing intensity are present at higher
frequencies, at energy intervals $\omega_0$. The satellites are also known as Frank-Condon 
side-bands~\cite{Allen99}, and their weight increases with increasing e-p coupling strength.
For larger coupling constants the {\it average} orbiton excitation energy increases,
caused by the increase of the crystal-field splitting, but low and high
energy satellites are always present due to the interaction of the orbital
excitation with lattice vibrations. 

The mixing of orbital and phonon mode gives rise to one extra phonon satellite
in the orbiton spectral function, at frequencies below $\omega_0$. 
In Fig.~\ref{Fig:local}b we see that, vice versa, due to the mixing 
a low intensity orbital satellite at $\approx 3J$ is present in the $Q_2$ phonon spectral function.
The $Q_2$ vibrational mode softens with increasing $g$, in contrast to the 
the $Q_3$ mode, which is not affected by the e-p interaction.

\begin{figure}
\epsfxsize=80mm
\centerline{\epsffile{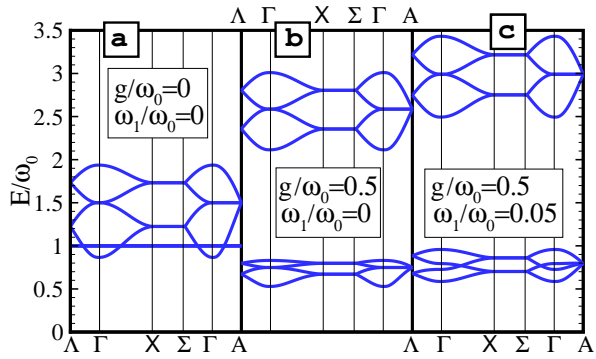}}
\vspace*{2mm}
\caption{
Orbiton and phonon dispersion, neglecting dynamical effects due to the e-p coupling;
(a) without e-p coupling $g$ and without bare phonon dispersion, (b) $g/\omega_0=1/2$, no bare phonon 
dispersion, and (c) $g/\omega_0=1/2$, finite bare phonon dispersion. The points of high symmetry 
in the Brillouin zone correspond to those of Ref.~[13].
}
\label{Fig:disp}
\end{figure}

{\it Full Hamiltonian.} 
The quadratic parts of the Hamiltonian in Eqs.(\ref{H_orb}-\ref{H_ph}) can be
diagonalized by a generalized Boguliobov transformation. For the moment we do not consider
the cubic terms, that give rise to the dynamical screening of the orbital excitation.
In the transformed operators the Hamiltonian is 
$
H_{quad}
=\sum_{\nu k} \epsilon_{\nu k} \alpha^{\dagger}_{\nu k}  \alpha_{\nu k},  
$
with
$
q^{\dagger}_{\nu k} = \sum_{\mu} 
u^k_{\nu \mu} \alpha^{\dagger}_{\mu k} + 
v^k_{\nu \mu} \alpha_{\mu -k}.
$
The energies of the eigenstates are given by
\begin{eqnarray}
\epsilon_{1,2 k}^2 &=& 
\zeta_k + \chi_k
\pm \left[ ( \zeta_k - \chi_k )^2  + 4 g^2 \bar{J} \omega_0 
\right]^{\frac{1}{2}} \nonumber \\
\epsilon_{3 k}^2 &=& \omega_0 ( \omega_0- 6 \gamma_{\parallel k}  
\omega_1 ),
\end{eqnarray}
with
$
\zeta_k = \bar{J}(\bar{J} - \gamma_{1k})/2
$
and
$
\chi_k = \omega_0 (\omega_0 + 2\gamma_{2k} \omega_1 )/2. 
$
The analytical expressions for $u^k_{\nu \mu}$ and $v^k_{\nu \mu}$ are
rather involved.
In Fig.~\ref{Fig:disp} the dispersion of the eigenmodes $\epsilon_{1 k}$ and
$\epsilon_{2 k}$ are shown for different sets of parameters, where we 
used $J_c/J=0.5$~\cite{Saitoh01}.
The modes with predominantly phonon character are centered around $E \approx \omega_0$,
and with orbital character around the crystal-field energy $\bar{J}$. For the parameters indicated in 
of Fig.~\ref{Fig:disp}b the wavefunctions of the low energy
excitations have on average 95\% phonon character and the high energy modes 95\% orbital character. 
Note that the orbiton dispersion is almost entirely due to the exchange coupling $J$ and
that via the e-p coupling the orbiton dispersion reflects itself in the effective phonon dispersion.

\begin{figure}
\epsfxsize=80mm
\centerline{\epsffile{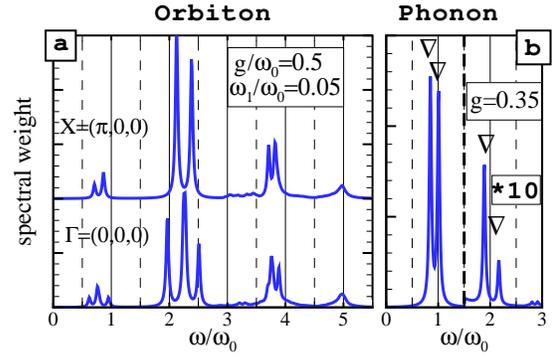}}
\vspace*{2mm}
\caption{
(a) Orbiton spectral function at the $\Gamma$ and X-point, $g=\omega_0/2$. 
(b) Spectrum of the Raman-active A$_{\rm g}$ and B$_{\rm 1g}$ phonon modes for
$\omega_1 / \omega_0=0.05$ and $g/ \omega_0=0.35$.
The experimental peak positions are indicated by $\nabla$. For $\omega > \omega_0$
the spectral weight is multiplied by 10, $g=0.35 \omega_0$, $\omega_1 / \omega_0=0.05$.
}
\label{Fig:spec}
\end{figure}

We now consider the interaction between the eigenmodes via the cubic term in 
Eq.(\ref{H_e-p}). 
The cubic terms can be taken into account
in a diagrammatic expansion. The first non-zero diagram
corresponds to an orbiton that excites a phonon, propagates and absorbs the phonon again, 
which is a second order process. 
We calculate the self-energy due to this process self-consistently,
i.e. instead of the bare orbiton propagator, we use the orbiton propagator dressed
with phonon excitations.
This is equivalent to the self-consistent Born approximation,
as is used, for instance, to calculate properties of a single hole in the t-J like 
models~\cite{BHO00,Kane89}. 
This approximation
works exceptionally well in t-J like models, for the problem of a single hole that is strongly 
coupled to magnons, or phonons~\cite{Ramsak92}, 
giving us confidence in its accuracy for the problem of a single orbital 
excitation coupled to phonons, which we consider here.

After the Boguliobov transformation, the single cubic term in Eq.(\ref{H_e-p}) maps into eight
different non-zero cubic combinations of the new $\alpha_{\nu k}$ operators. 
We can make use of the observation that,
in the parameter regime that we consider, the eigenmodes $\epsilon_{1,2 k}$ have almost entirely
orbiton c.q. phonon character. The calculated self-energy is therefore dominated by only one
of the eight cubic combinations of $\alpha_{\nu k}$ operators, which is at least an order of 
magnitude larger than other terms. Physically, this term is due to the orbiton-phonon
scattering process described in the previous paragraph. Details of the derivation will be published elsewhere~\cite{ToBePub}. 

The self-consistent calculation is performed numerically, taking
about $10^4$ points in the Brillouin-zone and an energy-grid with meshsize
$\omega_0/100$. The resulting orbiton spectral function at two high symmetry 
points in the Brillouin zone 
is shown in Fig.~\ref{Fig:spec}a, where we used the same parameters as in 
Fig.~\ref{Fig:disp}c. Comparing these two figures, we see that due to the dynamical e-p
coupling the poles with mainly orbital character are shifted to lower frequency and that
at higher frequency phonon induced satellites develop. This is not unexpected, as the
same happens for the system in the localized limit (see Fig.~\ref{Fig:local}a). A closer look
to the orbital spectrum at the $\Gamma$-point, however, reveals also that the effective
{\it orbital dispersion} is $\omega_0/2$, whereas the free orbital dispersion is $\omega_0$.
The reduced dispersion can be understood as a consequence of polaronic band-narrowing: the
effective mass of the orbital excitation increases because of its dressing with
phonons.

Finally we can compare the calculated spectral function of the Raman-active 
A$_{\rm g}$ and B$_{\rm 1g}$ phonon modes, shown in Fig.~\ref{Fig:spec}b, with 
experiment~\cite{Saitoh01}. 
The main phonon lines below $\omega_0$ and the weak orbiton induced satellites at
$\approx 2\omega_0$ are in excellent agreement with experiment. The orbiton satellites
are, just as in the localized limit (see Fig.~\ref{Fig:local}b), due to the mixing of
orbital and phonon modes.
The value of the e-p coupling that is used in the phonon calculation, $g/ \omega_0=0.35$, 
corresponds to rather weak
electron-phonon coupling, in accordance with Ref.~\cite{Benedetti99}, but in contrast to
Ref.~\cite{Allen99}. If we were to use a larger value of $g$, the
orbiton dispersion would become too small.
A way to determine the e-p coupling regime experimentally is
to check for additional satellites in the Raman spectrum at about $\omega \approx 3 \omega_0$,
which for the coupling strength in the present calculation have low intensity, but would 
have large intensity if the system were in the strong coupling regime~\cite{Allen99}. 
It is crucial that we took the phonon-dynamics into account, as 
a purely static Jahn-Teller distortion (as in Ref.~\cite{Saitoh01}, see
also Fig.~\ref{Fig:disp}c) 
would lead to an orbital dispersion that is about a factor two too large.
We interprete the Raman peaks around 150 meV as orbiton satellites of the
phonon peaks around 80 meV, which would disappear without electron-phonon coupling. This is in contrast to Saitoh et al.~\cite{Saitoh01}, 
where the peaks are
due to different Raman scattering mechanisms and are assumed to be
independent, which implies strong dependence of their intensity ratios on 
incident photon energy.

{\it Conclusions.}
We calculated the orbiton and phonon properties for a realistic model Hamiltonian for LaMnO$_3$ 
and compare the results with Raman-scattering data.
We treat electron-electron correlations and lattice dynamics on equal footing and have shown 
that the orbiton dispersion, which is mainly caused by correlation effects, is strongly 
reduced by the electron-phonon coupling. This coupling also mixes the orbiton and phonon
modes and causes satellite structures in the orbiton and phonon spectral function. 
The elementary excitations of the system, in other words, are mixed modes with both orbital 
and phonon character. 
This leads us to interprete the features around 150 meV in recent 
Raman-scattering 
experiments on LaMnO$_3$ as orbiton derived satellites in the phonon spectral function.
These satellites should also be observable in other experiments that probe
phonon dynamics, for instance in neutron scattering.



\begin{references}
\bibitem{mixed_terms}
Note that the 'mixed terms' in $H_{orb}$, containing products 
$\tau^x \tau^z$, are neglected. This is a 
reasonable approximation~\cite{BHMO99,BHO00}. This also holds for
terms of the same symmetry in $H_{ph}$.
\end{references}
\end{document}